\documentclass[epjST]{svjour}

\usepackage{graphicx}
\usepackage[intlimits]{amsmath}
\usepackage{amssymb}
\usepackage{amsfonts}
\usepackage{bm}

\begin{document}
\title{Collective motion of active Brownian particles in one dimension}
\author{Pawel Romanczuk\inst{1}\fnmsep\thanks{\email{romanczuk@physik.hu-berlin.de}} \and  Udo Erdmann\inst{2} }
\institute{Department of Physics, Humboldt Universit{\"a}t zu Berlin, Newtonstr. 15, 12489 Berlin \and Helmholtz-Gemeinschaft, Anna-Louisa-Karsch-Stra\ss e 2, 10178 Berlin}
\abstract{
We analyze a model of active Brownian particles with non-linear friction and velocity coupling in one spatial dimension. The model exhibits two  
modes of motion observed in biological swarms: A disordered phase with 
vanishing mean velocity and an ordered phase with finite mean velocity. Starting from the microscopic Langevin equations, we derive mean-field equations of the collective dynamics. We identify the fixed points of the mean-field equations corresponding to the two modes and analyze 
their stability with respect to the model  parameters. Finally, we compare our 
analytical findings with numerical simulations of the microscopic model.
} 
\maketitle
\section{Introduction}
\label{intro}
The concept of active Brownian particles was introduced more than a decade ago \cite{schweitzer_clustering_1994,ebeling_active_1999,schweitzer_brownian_2003} and was applied to different problems such as for example structure formation in excitable and chemotactic systems \cite{schimansky-geier_structure_1995,romanczuk_beyond_2008} or diffusion and swarming in biological systems \cite{schimansky-geier_advantages_2005,strefler_swarming_2008}. A general property of an active system is the emergence of complex behavior in the absence of external forces, which may be associated with internal degrees of freedom. 
On the one hand, there are many biological systems which can be interpreted as active particles ranging from individual animals \cite{komin_random_2004} to molecular motors \cite{lindner_some_2009}. On the other hand, recent reports of artificial systems far from equilibrium show characteristics of active Brownian motion \cite{liehr_drift_2003,sumino_self-running_2005,ruckner_chemically_2007} and demonstrate the general applicability of the concept to a wide range of natural phenomena.

In this work, we will analyze the collective motion (swarming) of active Brownian particles with a velocity alignment interaction. Collective motion of living organisms is a fascinating self-organization phenomenon which has attracted scientists from various disciplines \cite{parrish_complexity_1999,couzin_collective_2002,paley_oscillator_2007} and has led also to numerous publication in the field of statistical physics, nonlinear dynamics and pattern formation \cite{vicsek_novel_1995,toner_flocks_1998,czirok_collective_1999,simha_hydrodynamic_2002,romanczuk_collective_2009,ginelli_large-scale_2010}. Whereas most studies concentrates on simple models of collective motion of self-propelled particles with constant speed, only few publications consider collective motion of (active Brownian) particles with non-trivial speed dynamics \cite{niwa_self-organizing_1994,dorsogna_self-propelled_2006,ebeling_swarm_2008,strefler_swarming_2008,strefler_dynamics_2009}. In this work we derive in a systematic way mean-field equations for collective motion of active Brownian particles. We discuss the analytic results and compare the prediction of the mean field theory with numerical simulations.       

\section{Microscopic Model}
\label{sec:micro}
We consider a one-dimensional system of $N$ active Brownian particles with mass $m=1$. The evolution of the particle positions $x_i$ and velocities $v_i$ is described by the following set of (stochastic) differential equations ($i=1\dots N$):
\begin{align}
\dot x_i & =  v_i \\
m\dot {v}_i & = (\alpha-\beta{v}^2_i){v}_i + \mu (u_{\varepsilon,i} - {v}_i) +\sqrt{2 D} {\xi}_i \label{eq:langevin}
\end{align}
The first term on the right hand side is the non-linear friction force, the so-called Rayleigh-Helmholtz friction, of active Brownian motion, studied in \cite{erdmann_brownian_2000,erdmann_excitation_2002,lindner_diffusion_2008}. The second term describes a velocity alignment interaction with 
\begin{align}
u_{\varepsilon,i}= \frac{1}{N_\varepsilon} \sum_{\substack{j=1 \\ j \neq i}}^{N_\varepsilon} {v}_j,
\end{align}
i.e. with the mean velocity of particles within a finite distance $|x_j-x_i|<\varepsilon$ around the focal particle $i$ ($N_\varepsilon$ - number of neighbors within $\varepsilon$); $\mu$ denotes the alignment strength, which is positive $\mu\geq0$. Finally, the last term on the right hand side of Eq.~(\ref{eq:langevin}) is a white Gaussian noise with intensity $D$ and  $\langle \xi_i\rangle =0$, $\langle \xi_i(t)\xi_j(t')\rangle=\delta_{ij}\delta(t-t')$.   

For a large number of neighbors $N_\varepsilon\gg 1$ moving with random velocities (disordered state), the mean velocity of neighboring particles vanishes $u_{\varepsilon,i}=0$. In this situation, the mean squared stationary velocity of a single particle is given as $v_{0}^2=(\alpha-\mu)/\beta$. For perfectly ordered motion where all particles move with the same velocity ($u_{\varepsilon,i}=v_i$, $D=0$), the alignment force vanishes and we obtain $v_{0}^2=\alpha/\beta$.

\section{Mean Field Theory}
\label{sec:meanfield}
We derive a mean field transport theory of a gas of active Brownian particles based on the formulation of moment equations for the particle probability distribution. In general, for a system far from equilibrium the probability distribution is not Gaussian, and a correct description requires infinitely many moments (see for example \cite{pawula_approximation_1967,pawula_approximating_1987}). Here we perform a closure of our moment equations by neglecting temperature fluctuations $\theta$ (Eqs.~(\ref{eq:m4}),~(\ref{eq:theta})). 
The approximation of a non-Gaussian probability distribution by a finite number of moments may lead to unphysical behavior, such as negative values or artificial oscillations of the (approximated) probability distribution. Therefore, we will later compare the analytical results to numerical simulations of the microscopic system.

The $n$-th moments of the velocity $ \langle v^n \rangle $ is defined as
\begin{equation}
M_{n}(x,t) = \langle v^n \rangle = \frac{1}{\rho} \int v^n P(x,v,t) dv, \quad n > 0,
\label{eq:moments}
\end{equation}
where $P(x,v,t)$ is the probability density function describing the probability to find a particle at time $t$, at position $x$ moving with velocity $v$. The normalization $\rho$ is the zeroth moment which is equivalent to the marginal density
\begin{equation}
M_{0}(x,t) = \rho(x,t) = \int P(x,v,t) dv.
\end{equation}
Multiplying the $n$-th moment with the density and taking the derivative with respect to time, we obtain the dynamics of the moments of velocity
\begin{equation}
\frac{\partial}{\partial t}(M_{0} M_{n}) = \int v^n \frac{\partial P}{\partial t} dv.
\label{eq:n-moment}
\end{equation}
We start with an effective single particle description and omit for simplicity the individual particle index $i$. The Fokker-Planck-Equation for a single particle in the mean velocity field $u_\varepsilon$ of others reads
\begin{align}\label{eq:fpe}
\frac{\partial P}{\partial t} &  = - v \frac{\partial}{\partial x} P - \frac{\partial}{\partial v} \lbrace (\alpha - \beta v^2)v + \mu(u_{\varepsilon}-v)\rbrace P + D  \frac{\partial^2}{\partial v^2} P  . 
\end{align}
Inserting  Eq.~(\ref{eq:fpe}) in Eq.~(\ref{eq:n-moment}) and using $\lim_{v\to\pm\infty} P(x,v,t)=0$, the terms with partial derivatives with respect to $v$ can be partially integrated, yielding 
\begin{align}\label{eq:dynmoment}
\frac{\partial}{\partial t}(M_{0} M_{n})   = &  - \frac{\partial}{\partial x} \rho \ \langle v^{n+1} \rangle  
						+ n \ \rho \left[ \alpha \ \langle v^n \rangle - \beta \  \langle v^{n+2} \rangle  
						+ \mu(u_\varepsilon\langle v^{n-1}\rangle - \langle v^{n}\rangle)\right] \nonumber \\
					     & + n  \left(n-1\right) \ D \ \rho \ \langle v^{n-2} \rangle.
\end{align}
We rewrite velocity of the focal particle as a sum of the local velocity field $u(x,t)$ plus some deviation $\delta{v}$: $v=u+\delta v$. Furthermore, we assume $\langle \delta v^l \rangle=0$ for odd exponents $l$ ($l=1,3,5,\dots$). Thus we obtain for the moments (up to $l=4$):
\begin{subequations}\label{eq:m}
\begin{align}
\langle v \rangle \, &= u \label{eq:m1} \ ,\\
\langle v^2 \rangle &= u^2 + T \label{eq:m2}\ ,\\
\langle v^3 \rangle &= u^3 + 3 \ u \ T  \label{eq:m3}\ ,\\
\langle v^4 \rangle &= u^4 + 6 \ u^2 \ T + T^2 + \theta \label{eq:m4}\ .
\end{align}
\end{subequations}
Here, $T$ is the mean squared velocity deviation $T=\langle \delta v^2 \rangle$, which we will refer to as the temperature of the active particle gas, whereas $\theta$ is the average of the mean squared temperature fluctuations defined as 
\begin{align}\label{eq:theta}
\theta = \langle \left(\left(v - u\right)^2 - T\right)^2 \rangle= \langle \delta v^4 \rangle - T^2.
\end{align}
Now we can insert the Eqs.~(\ref{eq:m}) in Eq.~(\ref{eq:dynmoment}). Considering the dynamics up to $n=2$, after some calculus we arrive at a set of three coupled partial differential equations for the evolution of the density $\rho(x,v,t)$, the mean velocity field $u(x,v,t)$, and the temperature field $T(x,v,t)$:
\begin{subequations}
\begin{align}
 \frac{\partial}{\partial t} \rho  = &  -\frac{\partial}{\partial x} \left(\rho \ u\right) \\
 \frac{\partial u}{\partial t} + u \ \frac{\partial}{\partial x} u = & \ \alpha \ u - \beta \ u \ \left(u^2 + 3 T\right)+\mu(u_\varepsilon-u)
 - \frac{\partial T}{\partial x} - \frac{T}{\rho} \ \frac{\partial \rho}{\partial x} \\
 \frac{1}{2} \left(\frac{\partial T}{\partial t} + u \ \frac{\partial \ T}{\partial x}\right) = & \ (\alpha-\mu) \ T - \beta \ T \left(3 u^2 + T \right)  - \beta \ \theta  + D - T \frac{\partial u}{\partial x}\label{eq:mftemp} 
\end{align}
\label{eq:dgls}
\end{subequations}
Let us consider for simplicity an isotropic system with vanishing gradients in mean velocity $u$ and temperature $T$, which is a good approximation of our systems at high particle densities. In this case, the local velocity in the velocity alignment force equals the constant mean velocity field across the system $u_\varepsilon=u$, and  we end up with the following two ordinary differential equations for the temporal evolution of $u$ and $T$:
\begin{subequations}
\begin{align}
\frac{d u}{d t}  = &\alpha \ u - \beta \ u \ \left(u^2 + 3T\right)\\
\frac{1}{2}\frac{d  T}{d t} =& (\alpha-\mu) \ T - \beta \ T \ \left(3u^2 + T\right) - \beta \theta + D. 
\end{align}
\label{eq:mfT}
\end{subequations}
In order to obtain a closed system of equations, we neglect the temperature fluctuations by setting $\theta=0$ in Eq.~(\ref{eq:theta}), which is a reasonable assumptions at small noise intensities. Thus, the above differential equations constitute a two-dimensional dynamical system, with 6 fixed points  (stationary solutions) in the $(u,T)$ phase space, which can be analyzed by means of linear stability analysis.

The stationary solutions ($d u/d t=d T /d t =0$) for $u$ and $T$ read:
\begin{subequations}
\begin{align}
u_{1,2} & = 0, \\ 
T_{1,2} & = \frac{\alpha-\mu \pm \sqrt{(\alpha-\mu)^2 +  4 \beta D}}{2 \beta} \\
u_{3,4} & = \pm \frac{\sqrt{10 \alpha -3 \left(\mu-\sqrt{(2\alpha+\mu)^2 - 32 \beta D}\right)}}{4 \sqrt{\beta}} \\
T_{3}=T_{4} & = \frac{2\alpha + \mu - \sqrt{(2\alpha+\mu)^2 - 32 \beta D}}{16 \beta} \\
u_{5,6} & = \pm \frac{\sqrt{10 \alpha -3 \left(\mu+\sqrt{(2\alpha+\mu)^2 - 32 \beta D}\right)}}{4 \sqrt{\beta}} \\
T_{5}=T_{6} & = \frac{2\alpha + \mu + \sqrt{(2\alpha+\mu)^2 - 32 \beta D}}{16 \beta}.  
\end{align}
\end{subequations}
The kinetic temperature $T_j$ has to be positive, therefore, for $u=0$, $T_1$ (positive square root) is the only physically reasonable solution. 

The first solution with vanishing mean velocity $u=0$ describes a disordered phase. For $D=0$, the temperature $T = \frac{\alpha-\mu}{\beta} = v_0^2$ equals the square of the stationary velocity of individual particles. The kinetic energy of all particles completely consists  of fluctuations, no systematic translational motion occurs.

\begin{figure}
\begin{center}
  \includegraphics[width=\linewidth]{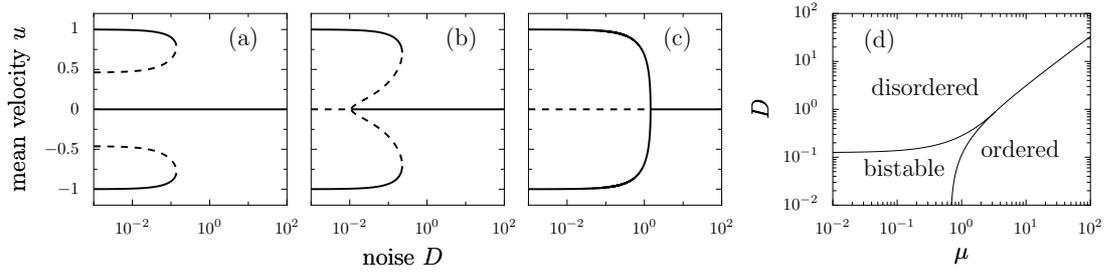} 
\end{center}
\caption{Bifurcation diagram of the mean velocity $u$ vs noise intensity $D$ as predicted from the mean field theory for different velocity alignment strengths ($\alpha=\beta=1$) : {\bf (a)} $\mu<2\alpha/3$ ($\mu=0.1$), {\bf (b)} $2\alpha/3<\mu<10\alpha/3$ ($\mu=0.7$) and {\bf (c)} $\mu>10\alpha/3$ ($\mu=5.0$). {\bf (d)} Phase diagram with respect to velocity alignment $\mu$ and noise intensity $D$.} 
\label{fig:ubif}       
\end{figure}
The second pair of solutions corresponds to translational modes which are stable below a critical noise intensity. The two solutions correspond to translational motion with positive or negative velocity $u$, thus, to a collective motion of the particles to the left or right. Without noise, $T_{3,4} = 0$, and the stationary mean velocity reduces to $u_{3,4} = \pm \sqrt{\alpha/\beta}$. Increasing the noise rises the kinetic temperature, and results in a decrease of the mean speed $|u|$.\\
The last solution pair describes unstable modes, for which with increasing noise intensity $D$ the temperature decreases and the mean speed increases. 

For low velocity alignment strength $\mu<2\alpha/3$, the disordered phase is always a stable solution; for  $\mu>2\alpha/3$, the linear stability analysis of the mean-field equations predicts the existence of a critical noise intensity 
\begin{equation}
D_{1, \rm crit} = \frac{\alpha(3\mu-2\alpha)}{9 \beta},
\label{eq:D_1}
\end{equation}
which determines the stability of the disordered solution. Starting from large noise intensities where the disordered solution is stable and decreasing the noise below $D_{1, \rm crit}$, we observe a pitchfork-bifurcation, and the disordered phase becomes unstable. Depending on the value of $\mu$, the pitchfork-bifurcation is either sub- or super-critical. For $\mu<10\alpha/3$, the disordered solution becomes unstable through a collision with the two unstable translational solutions, whereas for $\mu>10\alpha/3$ no unstable translational solutions exist and the disordered solution becomes unstable directly through the appearance of the two stable translational solutions (see Fig.~\ref{fig:ubif}). Thus, for $\mu>2\alpha/3$ and $D<D_{1,\rm crit}$, only the translational solutions $u_{3,4}$ are stable. 

\begin{figure}
\begin{center}
  \includegraphics[width=0.75\linewidth]{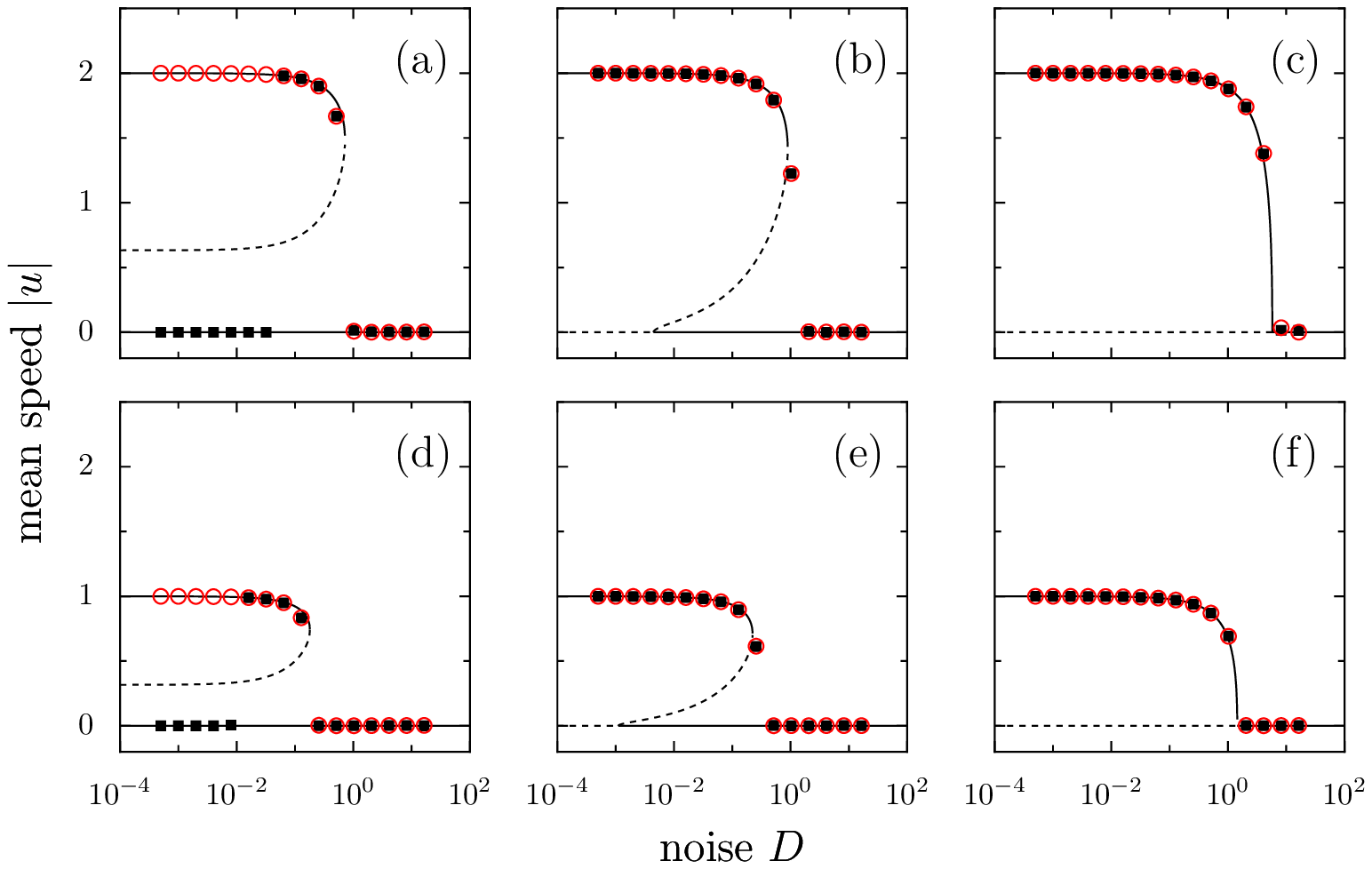}
\end{center}
\caption{(Color online) Comparison of the stationary solution obtained from simulation with theoretical prediction from the mean field theory for different velocity alignment strengths: $\mu=0.4$ {\bf (a,d)}, $\mu=0.67$ {\bf (b,e)} and $\mu=5.0$ {\bf (c,f)} and different friction coefficients: $\beta=0.25$ {\bf (a,b,c)} and $\beta=1.0$ {\bf (d,e,f)}. Other parameters used: particle number $N=8192$, simulation domain $L=500$, velocity alignment range $\epsilon=50$ and $\alpha=1.0$.  The initial conditions were either the disordered state (black filled squares) or the ordered state (red circles). Solid (dashed) lines show the stable (unstable) stationary solutions of the mean-field equations.}
\label{fig:usim}       
\end{figure}
For $\mu<10\alpha/3$ there exists a second critical noise intensity which determines the stability of the ordered phase (translational solutions, $u\neq0$). Above the critical noise intensity 
\begin{equation}
D_{2, \rm crit} = \frac{(2\alpha+\mu)^2}{32 \beta}
\label{eq:D_2}
\end{equation}
all translational solutions become unstable through a saddle-node bifurcation (Fig. \ref{fig:ubif} a,b).  

%
In order to test our analytical results, we performed numerical simulations of the microscopic model with periodic boundary condition. Due to the symmetry of the translational solutions $u_3=-u_4$, we distinguish the disordered phase and the ordered (translational) phase by measuring the global mean speed in our simulations:
\begin{align}
\left\langle |u| \right\rangle& =\left\langle \left| \frac{1}{N} \sum_{i=1}^{N} v_i  \right| \right\rangle.
\end{align}
Here, $\langle\cdot\rangle$ denotes temporal average after the system has reached a stationary state.
In order to analyze the stability of the (dis)ordered phase, the simulations were performed with two different initial states: perfectly ordered with $u(t=0)=u_3(D=0)$ and perfectly disordered state with $u(t=0)=0$. Each simulation was run until the system reached a stationary state but at least for $t=2000$ time units with a numerical time step $\Delta t=0.01$.  

\begin{figure}
\begin{center}
  \includegraphics[width=0.7\linewidth]{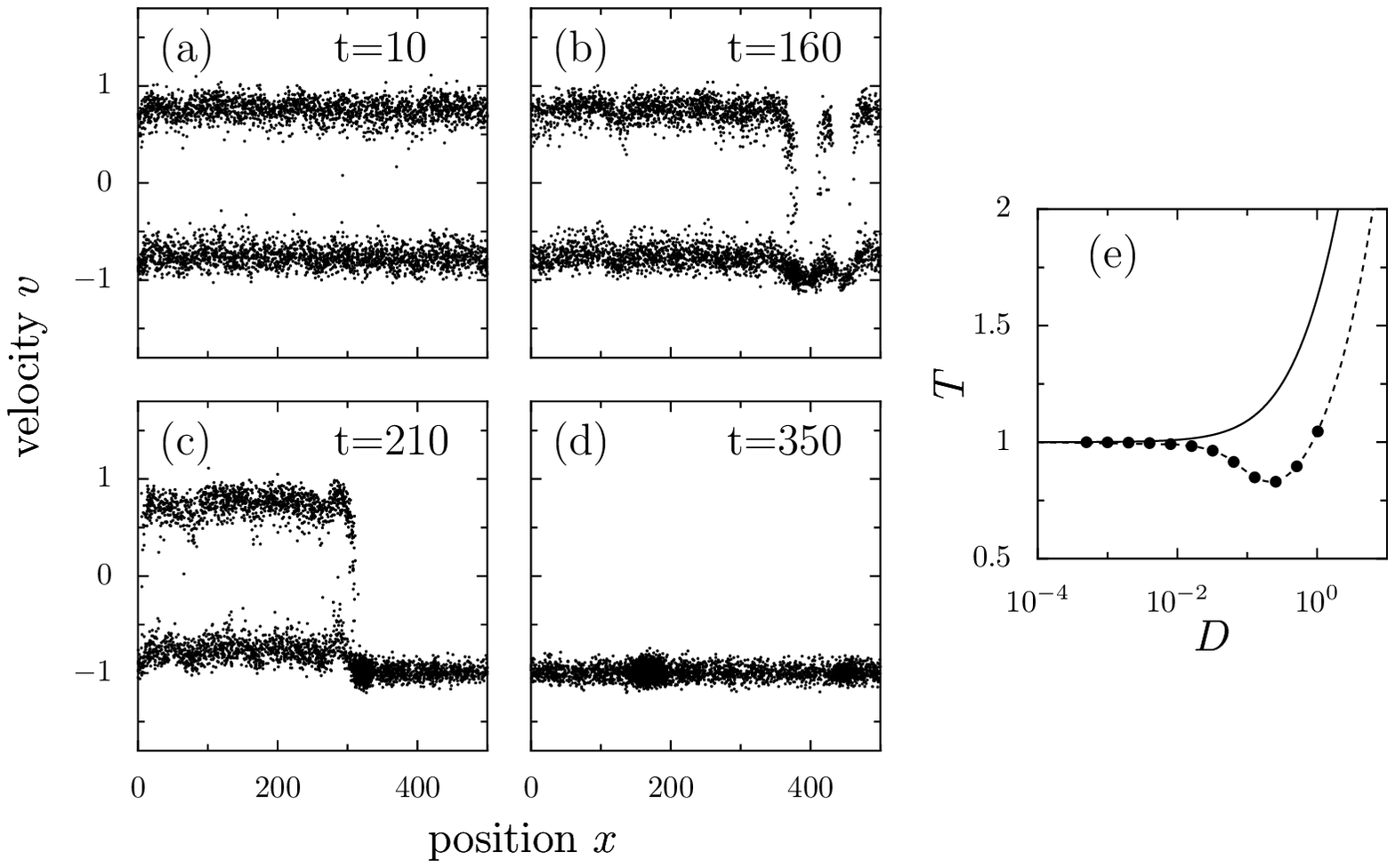} 
\end{center}
\caption{{\bf (a-d)} Simulation snapshots showing the breakdown of the disordered solution. Each point corresponds to a single particle. Starting from a perfectly disordered initial state ({\bf (a)}, $t=10$), we observe a local symmetry breaking at $x\approx400$  ({\bf (b)}, $t=160$) leading to a local velocity alignment of particles. The region of alignment spreads through the system ({\bf (c)}, $t=210$) until a stationary ordered state is reached ({\bf (d)}, $t=350$); Simulation parameters: $\alpha=1.0$, $\beta=1.0$, $\mu=0.4$, $D=0.01$, $N=4096$, $L=500$, $\epsilon=10$. {\bf (e)} Stationary temperature for non-interacting particles ($\mu=0$) vs noise intensity $D$ obtained from the mean field theory (solid line), from direct calculation of the second moment Eq.~(\ref{eq:rhm2}) (dashed line) and from numerical simulations (symbols) for $\alpha=\beta=1$.}
\label{fig:snapshots}       
\end{figure}

The stationary speed of the ordered phase vs noise intensity obtained from numerical simulations with ordered state initial condition are in a good agreement with the theoretical predictions from the mean field theory. Whereas simulations with disordered initial condition reveal an unexpected instability of the disordered solution. At finite $\mu$ the numerical simulations show that at intermediate $D$ the disordered solution $u=0$ becomes unstable via a spontaneous symmetry breaking as shown in Figs. \ref{fig:usim} and \ref{fig:snapshots}a-d, which is not predicted by the mean field theory.  

\section{Discussion and Summary}
We confirmed the instability of the disordered solution in numerical simulations for intermediate noise strengths for different particle numbers ($N=4096,8192,16384$) and obtained the same result for all values of $N$. This suggests that it cannot be simply dismissed as a pure finite size effect. Other possible sources for this discrepancy might be the restriction to spatially homogeneous solutions of the mean-field equations, which neglects density fluctuations, or the closure of the moment equation hierarchy by neglecting temperature fluctuations $\theta$. The latter is consistent with the observation that the deviation does not appear at low $D$, where temperature fluctuations are very small. In fact, for non-interacting particles with Rayleigh-Helmholtz friction we can calculate directly the second moment from the velocity distribution \cite{erdmann_brownian_2000}. For $\beta=1$ it reads  
\begin{align}\label{eq:rhm2}
\langle v^2 \rangle = \frac{\alpha}{2}\left(1+
			\frac{\pi \sqrt{2} I_{-\frac{3}{4}}\left(\frac{\alpha^2}{8 D}\right)  - K_{-\frac{3}{4}}\left(\frac{\alpha^2}{8 D}\right)}{K_{\frac{1}{4}}\left(\frac{\alpha^2}{8 D}\right)+\pi\sqrt{2}I_{\frac{1}{4}}\left(\frac{\alpha^2}{8 D}\right)}\right),
\end{align}
with $I_n$ and $K_n$ being the modified Bessel functions of the first and second kind respectively. The result of Eq.~(\ref{eq:rhm2}) corresponds directly to the temperature $T$ for the disordered state in the limit $\mu=0$. Due to the non-linearity of the friction function, the temperature does not increase monotonically with $D$ as predicted by the mean field theory but exhibits a minimum at intermediate noise intensities as shown in Fig. \ref{fig:snapshots}e. It can be seen from Eq.~(\ref{eq:mfT}) that the explicit consideration of finite temperature fluctuations ($\theta>0$) leads to a decrease in $T$, which is consistent with Eq.~(\ref{eq:rhm2}). This in turn decreases the stability of the disordered state. The extension of the mean field theory to higher orders would account for this effect at the expense of the analytical tractability of the mean field solutions.     

In summary, we have formulated a one-dimensional model of active Brownian particles with non-linear Rayleigh-Helmholtz friction interacting through a local velocity alignment. Starting from the microscopic Langevin description, we derived a set of mean-field equations via the corresponding Fokker-Planck Equation. By neglecting temperature fluctuations, we obtained a set of coupled partial differential equations for the density, velocity and temperature fields. For simplicity, we restricted the analysis here to the spatially homogeneous situation. We have identified mean-field solutions corresponding to an ordered mode (collective motion) and to the disordered mode with random particle velocities and analyzed their stability. A comparison of our analytical results with numerical simulations of the microscopic model confirms our findings for the ordered phase, but shows deviations with respect to the stability of the disordered solutions.     

P. R. would like to thank M. Kostur for the introduction to numerical simulations on Graphical Processing Units using CUDA which allow a considerable speed up of numerical simulations \cite{januszewski_accelerating_2010}. U. E. would like to thank B. Eckehardt for fruitfull discussions on active Brownian particles. Last but not least, we would like to express our gratitude to L. Schimansky-Geier for his scientific input in our research and the time we went together; we wish him all best in the future.   


\end{document}